\documentstyle[amstex,aps]{revtex}
\begin{document}
\title{Quntum Dynamics in Stochastic Mechanics}
\author{J. M. A. Figueiredo}
\address{Universidade Federal de Minas Gerais - Dept. de F\'{i}sica\\
Caixa Postal 702 - Belo Horizonte - Brazil - 30.123-970}
\date{\today}
\maketitle

\begin{abstract}
We present a study of the motion that a massive particle in a
non-dissipative Brownian motion in vacuum is subject. Noise source comes
from vacuum fluctuations of some quantum field capable of interact with the
particle. For the associated Fokker-Planck equation we do a perturbation
theory which has terms presenting dynamics similar to that satisfied by the
probability amplitude in Quantum Theory even though the physical
interpretation of these terms be classical. In particular observables
present expected values that coincide with those calculated using Quantum
Mechanics.
\end{abstract}


\newpage

\subsection{\protect\bigskip {\bf Introduction}}

The probabilistic character of \ Quantum Mechanics as well as the formal
similarity of the Schr\"{o}edinger equation to the diffusion equation raised
from long ago the question of a possible stochastic justification to Quantum
Mechanics \cite{nelson1}. A first formal approach was developed by Nelson 
\cite{nelson} and up to now many authors presented complementary or
alternative works (a few of them listed in \cite{olavo}-\cite{suares}) on
the subject named usually as Stochastic Mechanics. A common goal of these
theories lie on the possibility of find a stochastic process that is
equivalent to Quantum Mechanics. Nevertheless no such a process was yet
proposed that lent to Stochastic Mechanics, in any of its presented forms, a
explicit advantage over the sophisticated epistemological structure of
Quantum Theory. On another way Quantum Theory is continuously tested at
deeper levels fixing its place as fundamental one although this do not
invalidate the search for alternative interpretations, hopefully not subject
to inconsistencies (like infinities) yet plaguing the theory. In this work
we focus on a somewhat different approach where a particular kind of
stochastic process is studied. We present a perturbation theory for the
associated probability density function that has quantum formalism as a
subjacent dynamics but still preserving the classical nature of the theory.
We also discuss a possible physical origin for the noise source capable of
describe the type of process we study. No full equivalence to Quantum Theory
was found so this indicates that Quantum Mechanics as a fundamental theory
is preserved also. Nevertheless our formulation shows explicitly that
quantal aspects may appear quite naturally in a classical theory.

Once every physical process include or can generate some sort of noise the
motion of a massive particle under its influence is an important subject in
the study of real particle's dynamics. For a particular class of noise - the
Wiener process - a rigorous probability formalism can be obtained where a
field equation - the Fokker-Planck equation (FPE) - for the probability
density function of some noise-dependent variable is defined \cite{gardiner}%
. For this case noise source has an uncorrelated stationary probability
distribution and when the stochastic process under study describes
specifically the motion of a particle the resulting process is the Brownian
motion. Due to its generality a FPE can be considered as a fundamental
although its detailed predictions are dependent of the specific Wiener
process considered.

The origin of noise that a particle in Brownian motion is subject usually is
thermodynamic. This is the case of a particle dispersed on a fluid at some
temperature. There\ thermal noise gives a bit of energy to the particle and
dissipative effects returns it back (randomly) to the thermal bath. The net
effect is a random walk equilibrated to the bath's temperature. Here we will
consider this kind of motion but with two main differences. First the
particle is in vacuum and noise is generated by vacuum fluctuations of some
quantum field capable of interaction with that particle. Second the motion
is not dissipative. This last hypotesis seems to be at first sight
unrealistic because particle may gain energy from nothing. We accept it for
simplicity since this hypotesis do not invalidate the methods we present
here for the general case.

One may argue against the reality of vacuum fluctuations itself. This
subject has yet some controversy but its theoretical as well as experimental
grounds has been sufficiently discussed in the literature \cite{milonni} and
we can accept it in a general basis. Moreover recent careful measurements of
Casimir forces \cite{lamoreaux} demonstrated a good agreement between
observation and theoretical predictions reinforcing our belief in its
physical existence. Generally we even can believe that geometry changes in a
physically confined systems may induce a small violation of energy
conservation principle, undetectable by today's instrument sensitivity or be
possible that during molecular conformational changes some energy comes from
vacuum, involving van der Walls force. We think these arguments are
sufficient to justify a possible existence of a natural noise source
interacting to any elementary particle coming from vacuum properties.
Besides all we must also have in mind that every charged particle is subject
to the cosmic background radiation field that is real and also stochastic.
Therefore we think that stochastic process are inevitable to the motion of
massive particles in vacuum, coming as a intrinsic part of the reality in
Nature and must be considered in any precise description of this kind of
motion.

The method we give here to solve the FPE for the Brownian motion has general
validity and no fundamental difficulties are present when the more general
case where not only dissipative effects as well as a larger class of \
Wiener process are considered. Specifically for a quantum noise coming from
vacuum fluctuations we understand a stationary Wiener-type source with noise
intensity proportional to a \ ''vacuum power '' which we write as ${\cal P}%
\equiv 
%TCIMACRO{\UNICODE[m]{0xbc}}%
%BeginExpansion
{\frac14}%
%EndExpansion
\hbar \left\langle \omega ^{2}\right\rangle $ and with mean energy ${\cal E}%
\equiv 
%TCIMACRO{\UNICODE[m]{0xbd}}%
%BeginExpansion
{\frac12}%
%EndExpansion
\hbar \sqrt{\left\langle \omega ^{2}\right\rangle }$ and or generally 
\begin{equation}
{\cal P}\hbar ={\cal E}^{2}  \label{VP}
\end{equation}
Here $\hbar $\ is the Planck's constant and $\left\langle \omega
^{2}\right\rangle $\ is the variance of the field frequencies averaged over
some appropriate distribution (we assume $\left\langle \omega \right\rangle
=0$\ since $\omega $ and $-\omega $ must be considered as independent
fluctuations). For example in the cosmic background case where $T\simeq 2K$
we find ${\cal P}\simeq 1.15pW$.\ Calculation of $\left\langle \omega
^{2}\right\rangle $ for quantum fluctuations is not trivial because vacuum
energy density diverges as $\omega ^{3}$\ \cite{milonni} with (assumed)
uniform probability distribution denying a simple averaging process unless a
physical cutoff at high frequencies exist. This must be achieved by
multiplying the instantaneous vacuum modes by the appropriate cross section
(in that frequency) which falls off properly at high energies. If doing so
the instantaneous momentum transferred to the particle may be calculated
resulting in a noise source with zero mean and the desired power in as much
as in the Casimir effect \cite{shwinger}\ calculation where high frequencies
modes cancel out and the net effect comes from low frequencies modes only.
This is analogous to the thermal effect where the enormous kinetic energy of
one mol of molecules at room temperature has minimal action over the center
of mass motion of a macroscopic body because the random collisions between
the molecules and the body are isotropic so the net resulting effect are
small fluctuations. Nevertheless if the phenomena we describe here has to be
universal it is reasonable that ${\cal E}$\ should be a characteristic
mensurable constant in Nature even though we concentrate here only on the
effect on particle motion of a noise source with properties listed above and
explicit in the perturbative series expansion procedure we give below.
Therefore besides eqn $\left( \ref{VP}\right) $\ detailed physical origin of 
${\cal P}$\ (or ${\cal E}$) is not relevant to our calculations in the
present stage of development of our reasoning.

\bigskip

\subsection{\protect\bigskip {\bf Perturbation expansion for the
Fokker-Planck equation}}

In sequence we give the precise mathematical formulation of the kind of
motion we want to describe. Particle's dynamics arise from Hamilton
equations plus a Langevin term acting as a stochastic force. We assume no
explicit stochastic term for position variable meaning that any random
effect on it comes indirectly through coupling with the momentum term
resulting in a non-dissipative Brownian motion. More precisely we have 
\begin{mathletters}
\begin{align}
dx\left( t\right) & =\frac{\partial H}{\partial p}dt  \label{H1} \\
dp\left( t\right) & =-\frac{\partial H}{\partial x}dt+\sqrt{2m{\cal P}}%
dW\left( t\right)  \label{H2}
\end{align}
Here $m$\ is particle's mass and $dW\left( t\right) $ a stationary
Wiener-type stochastic variable with noise power ${\cal P}$\ given as above.
Rigorously speaking if the particle has a electric charge it radiates when
accelerated which is a kind of an intrinsic unremovable dissipative effect.
In this work we assume these effects as small thus leaving out any radiation
reaction terms and deserving for a future work a proper inclusion of
appropriate corrections.

The associated Fokker-Planck equation for the joint $\Phi \left(
x,p,t\right) $\ probability density distribution in phase space is given by 
\cite{gardiner} 
\end{mathletters}
\begin{equation}
\frac{\partial \Phi \left( x,p,t\right) }{\partial t}=-\frac{\partial H}{%
\partial p}\frac{\partial \Phi }{\partial x}+\frac{\partial H}{\partial x}%
\frac{\partial \Phi }{\partial p}+m{\cal P}\frac{\partial ^{2}\Phi }{%
\partial p^{2}}  \label{FP}
\end{equation}
which unless a missing dissipative term has the same structure of the
Kramers' equation \cite{kramer}, orginaly written to describe the motion of
\ a large mass particle in a fluid at thermal equilibrium with a bath in
such a way that noise comes from the temperature effect cited above. As
already discussed this motion is unstable against infinitesimal stochastic
perturbations. In fact even if the intensity $\sqrt{2m{\cal P}}$ of the
Langevin term goes to zero a smeared phase space still results. We observe
that in this case the above equation reduces to the Liouville equation which
admits a class of solutions concentrated along the deterministic
trajectories in phase space describing particle's classical motion. In fact
if $H_{cl}\left( t\right) \equiv \frac{P_{cl}\left( t\right) ^{2}}{2m}%
+V\left( x_{cl}\left( t\right) \right) $\ is the Hamiltonian of the
(deterministic) Hamilton equations one class of solutions for Liouville
equation is 
\begin{equation}
\Phi \left( x,p,t\right) =A\exp (-\beta \left| \frac{p^{2}-P_{cl}\left(
t\right) ^{2}}{2m}+V\left( x\right) -V\left( x_{cl}\left( t\right) \right)
\right| )  \label{liouv solut}
\end{equation}
We see that trajectories (in phase space) no longer exist any more:\
position and momentum variables decouple and information about the system is
supplied by an ensemble of states in thermal equilibrium at temperature $%
\beta ^{-1}$. This temperature effect is the only memory of the ghost
stochastic effect and is caused by the known singular small noise expansion
of the FPE \cite{gardiner}. Thus there is a big difference if we take
Hamilton equations without noise to get a full deterministic motion and a
FPE with null noise term. In the first case the topology of phase space as
represented by Poincare maps are sets of points even when the motion is
chaotic whereas a probability density function continuously fills phase
space in the another case.

The structure of the FPE allows easy calculation of some averages. For
example time derivative of the expected value of the Hamiltonian is given by 
\begin{align}
\frac{d}{dt}\langle H\rangle & =\int_{-\infty }^{\infty }dxdp\left[ \frac{%
p^{2}}{2m}+V\left( x,t\right) \right] \Phi \left( x,p,t\right)  \nonumber \\
\frac{d}{dt}\langle H\rangle & ={\cal P}-\langle \frac{\partial V}{\partial x%
}\frac{p}{m}\rangle +\langle \frac{\partial V}{\partial x}\frac{p}{m}\rangle
={\cal P}  \label{power}
\end{align}
and the expected value of the momentum changes at the rate 
\begin{equation}
\frac{d}{dt}\langle p\rangle =-\langle \frac{\partial V}{\partial x}\rangle
\equiv \langle f\rangle  \label{newton}
\end{equation}
following in mean the second Newton's law. Note that the mean exchanged
momentum with stochastic source is zero but a quadratic dispersion on
momentum exists being this consequence of the missing dissipative effects.
This dispersion increases with time meaning that, as anticipated above, the
particle gains some energy from the stochastic source . Position variable is
also affected although there is no noise source acting directly on it. Its
mean value and the associated fluctuations change at the rate 
\begin{align}
\frac{d}{dt}\langle x\rangle & =\langle \frac{p}{m}\rangle  \label{veloc} \\
\frac{d}{dt}\left( \langle x^{2}\rangle -\langle x\rangle ^{2}\right) & =%
\frac{2}{m}\left( \langle px\rangle -\langle p\rangle \langle x\rangle
\right)  \label{pos-fluct}
\end{align}
Thereby classical laws of motion are satisfied only in mean no matter the
value of the noise intensity as shown by eqns $\left( \ref{newton}\right) $%
and $\left( \ref{veloc}\right) $. We must conclude that a field dynamics for
the probability density should exist even for zero noise power and it will
be shown bellow it follows exactly quantum dynamics when our perturbative
schema is considered.

We develop in sequence the perturbative expansion. To do this we shall use $%
\hbar $ in order to Fourier Transform $\Phi \left( x,p,t\right) $ in
momentum to the adjoint space ${\cal Y}$ associated to this variable. In the
present formalism this space refer to wavelengths of a mode decomposition of
the probability density function. Differently from Quantum Theory here
position and momentum are not related by an unitary transformation between
adjoint spaces. This happens for momentum and wavelengths of \ waves of
probability density having momentum $p=\hbar y$. It is in this space the FPE
has a form suitable to the structure of the perturbative series we look for.
Nextly we write 
\begin{equation}
\Phi \left( x,p,t\right) =\frac{1}{2\pi }\int_{-\infty }^{\infty }\chi
\left( x,y,t\right) e^{i\frac{py}{\hbar }}dy  \label{Broglie}
\end{equation}
so the FPE reads 
\begin{equation}
\partial _{t}\chi \left( x,y,t\right) =\frac{\hbar }{im}\partial
_{y}\partial _{x}\chi -\frac{y}{i\hbar }\frac{\partial V}{\partial x}\chi -%
\frac{y^{2}}{\tau \lambda ^{2}}\chi  \label{FP2}
\end{equation}
where $\lambda \equiv {\cal E}/\hbar c$\ is the mean vacuum fluctuation
wavelength and $\tau $\ the particle's Compton length divided by the
velocity of light. We stress that phase space probability density $\Phi
\left( x,p,t\right) $\ is similar but has to be not confused with the Wigner
function \cite{wigner} in Quantum Theory. The former satisfies exactly the
Liouville equation for zero noise power. Wigner function satisfies it only
approximately. Moreover in a Fourier transform on momentum variable of the
Wigner function similar to eqn$\left( \ref{Broglie}\right) $ variable $y$\
has the meaning of position. Specifically in the Wigner function the role of 
$x$\ and $y$\ are symmetric so they share the same physical meaning whereas
in this work simbols\ $x$\ and $y$\ are trully indepent variables that play
different role. It will be shown explicitly later that in fact both
functions do not coincide.

Owing to the structure of the eqn$\left( \ref{FP2}\right) $\ it is natural
to think in a Taylor expansion of $\chi \left( x,y,t\right) $\ in the
variable $y$ which should be convenient because this way we may compare
probability wavelengths to vacuum wavelength (that is very big) as expressed
by the fraction present in the last term of this equation. This perturbative
schema has the advantage that it permits a controll of noise effects over
field dynamics for\ $\chi $. Saying differently if particle's kinetic energy
is much greater than vacuum energy ${\cal E}$\ its typical mode wavelength $%
y\sim \hbar /p$\ is very small compared to vacuum wavelength $\lambda $\
thus making noise effects negligible. But a infinite series in powers of $y$%
\ has no simple inverse Fourier representation\ generating a somewhat
pathological phase space reconstruction. For vanishingly small noise we know
from eqn$\left( \ref{liouv solut}\right) $\ that Liouville equation feels
the presence of a thermal bath with temperature $\beta ^{-1}$. In the
present case this effect is to be replaced by a ''vacuum temperature'' \
proportional to $\lambda ^{-1}$. Thus we insert a additional gaussian term
in the perturbation series in order to regularize phase space reconstrution
and to provide the appropriate thermal bath effect. More precisely we look
for a general series solution of the type 
\begin{equation}
\chi \left( x,y,t\right) =\exp \left( \frac{-y^{2}}{2\lambda ^{2}}\right)
\sum_{n=0}^{\infty }a_{n}\left( x,t\right) y^{n}  \label{series}
\end{equation}
in such a way that eqn$\left( \ref{FP2}\right) $\ becomes 
\begin{align*}
\sum_{n=0}^{\infty }\frac{\partial a_{n}\left( x,t\right) }{\partial t}%
y^{n}& =\frac{\hbar }{im}\sum_{n=0}^{\infty }\frac{\partial a_{n}\left(
x,t\right) }{\partial x}\left[ ny^{n-1}-\frac{y^{n+1}}{\lambda ^{2}}\right] -
\\
& \frac{1}{i\hbar }\frac{\partial V}{\partial x}\sum_{n=0}^{\infty
}a_{n}\left( x,t\right) y^{n+1}-\frac{1}{\tau \lambda ^{2}}%
\sum_{n=0}^{\infty }a_{n}\left( x,t\right) y^{n+2}
\end{align*}
generating the following set of differential equations for the coefficients $%
\left\{ a_{n}\right\} $ 
\begin{align}
\frac{\partial a_{0}\left( x,t\right) }{\partial t}& =\frac{\hbar }{im}\frac{%
\partial a_{1}\left( x,t\right) }{\partial x}  \label{a0} \\
\frac{\partial a_{1}\left( x,t\right) }{\partial t}& =\frac{\hbar }{im}\left[
2\frac{\partial a_{2}\left( x,t\right) }{\partial x}-\frac{1}{\lambda ^{2}}%
\frac{\partial a_{0}\left( x,t\right) }{\partial x}\right] -\frac{1}{i\hbar }%
\frac{\partial V}{\partial x}a_{0}\left( x,t\right)  \label{a1}
\end{align}
\begin{align}
\frac{\partial a_{n}\left( x,t\right) }{\partial t}& =\frac{\hbar }{im}\left[
\left( n+1\right) \frac{\partial a_{n+1}\left( x,t\right) }{\partial x}-%
\frac{1}{\lambda ^{2}}\frac{\partial a_{n-1}\left( x,t\right) }{\partial x}%
\right] -  \label{ann} \\
& \frac{1}{i\hbar }\frac{\partial V}{\partial x}a_{n-1}\left( x,t\right) -%
\frac{1}{\tau \lambda ^{2}}a_{n-2}\left( x,t\right)  \nonumber
\end{align}
$\qquad $

Before discussing solutions of the above set of equations note that a naive
interpretation can be given to some of these coefficients. In fact since 
\[
\chi \left( x,y,t\right) =\int_{-\infty }^{\infty }\Phi \left( x,p,t\right)
e^{-i\frac{py}{\hbar }}dp 
\]
we have 
\begin{align}
a_{0}\left( x,t\right) & =\chi \left( x,0,t\right) =\int_{-\infty }^{\infty
}\Phi \left( x,p,t\right) dp\equiv \left| \Psi \left( x,t\right) \right| ^{2}
\label{def a0} \\
a_{1}\left( x,t\right) & =\left( \frac{\partial \chi }{\partial y}\right)
_{y=0}=\frac{1}{i\hbar }\int_{-\infty }^{\infty }p\Phi \left( x,p,t\right)
dp\equiv \frac{1}{i\hbar }\pi \left( x,t\right)  \nonumber \\
a_{2}\left( x,t\right) & =\frac{1}{2}\left( \frac{\partial ^{2}\chi }{%
\partial y^{2}}\right) _{y=0}+\frac{a_{0}}{2\lambda ^{2}}=\frac{-1}{2\hbar
^{2}}\int_{-\infty }^{\infty }p^{2}\Phi \left( x,p,t\right) dp+\frac{a_{0}}{%
2\lambda ^{2}}  \label{a2}
\end{align}
that is $a_{0}$ is the marginal probability distribution, $a_{1}$\ is
proportional to the expected value of the momentum at point $x$\ and $a_{2}$%
\ is proportional to the kinetic energy at $x$. We have also introduced the
probability amplitude $\Psi $ belonging to the $L^{2}$ stuff of the $L^{1}$%
-based probability distribution $a_{0}$.\ Integrating eqns$\left( \ref{a0}%
\right) $-$\left( \ref{ann}\right) $ we get for $n=2$%
\begin{align*}
\frac{d}{dt}\int_{-\infty }^{\infty }\left| \Psi \left( x,t\right) \right|
^{2}dx& \equiv \frac{dN}{dt}=0 \\
\frac{d}{dt}\int_{-\infty }^{\infty }\pi \left( x,t\right) dx& \equiv \frac{%
d\left\langle p\right\rangle }{dt}=-\left\langle \frac{\partial V}{\partial x%
}a_{0}\left( x,t\right) \right\rangle =\left\langle f\left( x,t\right)
\right\rangle \\
-\frac{\hbar ^{2}}{m}\frac{d}{dt}\int_{-\infty }^{\infty }a_{2}\left(
x,t\right) dx& =\frac{1}{2m}\frac{d\left\langle p^{2}\right\rangle }{dt}=%
\frac{\left\langle f\left( x,t\right) \pi \left( x,t\right) \right\rangle }{m%
}+{\cal P}
\end{align*}
The first of these equations indicates that the norm of $\Psi $\ is
time-independent. Incidentally it also shows that the Hilbert space operator 
$i\hbar \frac{\partial }{\partial t}$ satisfy $\left( i\hbar \partial
_{t}\Psi ,\Psi \right) =\left( \Psi ,i\hbar \partial _{t}\Psi \right) $
where $\left( \cdot ,\cdot \right) $\ stands for Hilbert space inner
product. This does not mean at the present stage that $i\hbar \partial _{t}$%
\ is hermitian since the relation was not proved to be valid in general, for
any pair of Hilbert space vectors. However it is a necessary condition
suggesting that this operator admits a Hermitian representation on an
appropriate Hilbert space and in fact it will be constructed as the
consistence of our calcualtions becomes closed. The second equation is
similar to the Ehrenfest theorem and the third\ one indicates\ together with
eqn$\left( \ref{power}\right) $\ that the rise in total particle's energy
comes from its kinetic component. It's worthwhile note that the first two
equations are independent of noise intensity so concerning the
non-stochastic limit their validity has the same status as the Liouville
equation. In the next step we look for a field equation for $\Psi $. Define $%
L\equiv F\left[ P\right] +V\left( x,t\right) $, where $P\equiv \frac{\hbar }{%
i}\frac{\partial }{\partial x}$\ .We then have 
\[
\Psi ^{\ast }\left[ L,P\right] \Psi =-\frac{\hbar }{i}\frac{\partial V}{%
\partial x}\Psi ^{\ast }\Psi 
\]
and in consequence 
\begin{equation}
\frac{\partial }{\partial t}a_{1}\left( x,t\right) =\frac{\hbar }{im}\frac{%
\partial }{\partial x}\left[ 2a_{2}\left( x,t\right) -\frac{1}{\lambda ^{2}}%
a_{0}\left( x,t\right) \right] +\frac{\Psi ^{\ast }\left[ L,P\right] \Psi }{%
\hbar ^{2}}  \label{rate a11}
\end{equation}
\begin{equation}
\frac{d}{dt}\int_{-\infty }^{\infty }a_{1}\left( x,t\right) dx=\frac{1}{%
\hbar ^{2}}\left( \Psi ,\left[ L,P\right] \Psi \right)  \label{rate a1}
\end{equation}
\qquad Being $a_{1}$ pure imaginary we can write it as 
\begin{equation}
a_{1}=\frac{\Psi B\Psi ^{\ast }-\Psi ^{\ast }B\Psi }{2}  \label{def a1}
\end{equation}
\qquad \qquad where $B$ is necessarily a real-valued operator. Further
development on defining a more specific form for this operator also
generates the rules for the Hilbert space we are working on. Owing to eqn$%
\left( \ref{rate a1}\right) $ where a inner product emerges naturally we
choose $B$ such that $\widetilde{B}\equiv \frac{\hbar }{i}B$ is Hermitian
and time-independent giving the desired structure to expected value of the
momentum as shown below 
\[
\int_{-\infty }^{\infty }a_{1}\left( x,t\right) dx=\frac{1}{i\hbar }\left(
\Psi ,\widetilde{B}\Psi \right) \Rightarrow \left( \Psi ,\widetilde{B}\Psi
\right) =\langle \pi \rangle \equiv \Pi \left( t\right) 
\]
This result is to be interpreted classically even though it presents a
strong resemblance to quantum mechanical rules. The average force on the
particle is then equal to

\begin{equation}
\frac{d\Pi \left( t\right) }{dt}=\frac{d}{dt}\int_{-\infty }^{\infty
}a_{1}\left( x,t\right) dx=\frac{1}{\hbar ^{2}}\left[ \left( i\hbar \partial
_{t}\Psi ,\widetilde{B}\Psi \right) -\left( \Psi ,\widetilde{B}i\hbar
\partial _{t}\Psi \right) \right]  \label{force}
\end{equation}
and when compared with eqn$\left( \ref{rate a1}\right) $ suggest a simple
inner product structure for the above equation giving to $i\hbar \partial
_{t}$\ a possible Hermitian representation ${\cal H}$ which in general is
function of $x$, $P$, $\widetilde{B}$ and $t$. This option defines
simultaneously a field equation for the probability amplitude since now we
select those Hilbert space vectors that satisfy the equality 
\[
i\hbar \partial _{t}\Psi ={\cal H}\left( x,P,\widetilde{B},t\right) \Psi 
\]
When this equation is inserted in eqn$\left( \ref{force}\right) $\ and
compared to eqn$\left( \ref{rate a1}\right) $ we get

\[
\left( \Psi ,\left[ {\cal H},\widetilde{B}\right] \Psi \right) =\left( \Psi ,%
\left[ L,P\right] \Psi \right) 
\]
which should be valid for all $\Psi $\ that satisfy the above field
equation. So in an appropriate subspace we must have $\left[ {\cal H},%
\widetilde{B}\right] =\left[ L,P\right] $ with a formal solution given by 
\begin{eqnarray*}
\widetilde{B} &=&P \\
{\cal H}\Psi &=&L\Psi =\left[ F\left[ P\right] +V\left( x,t\right) \right]
\Psi
\end{eqnarray*}
which must be simultaneously valid. This way we conclude that $F\left[ P%
\right] $\ is Hermitian. Now eqn$\left( \ref{a0}\right) $ should also be
satisfied implying that 
\[
i\hbar \partial _{t}\left| \Psi \left( x,t\right) \right| ^{2}=\Psi ^{\ast }%
{\cal H}\Psi -\Psi {\cal H}\Psi ^{\ast }=\frac{\hbar ^{2}}{2m}\left( \Psi
\partial _{x}^{2}\Psi ^{\ast }-\Psi ^{\ast }\partial _{x}^{2}\Psi \right) 
\]
or 
\[
\Psi ^{-1}\left( F\left[ P\right] +\frac{\hbar ^{2}}{2m}\partial
_{x}^{2}\right) \Psi =\left( \Psi ^{\ast }\right) ^{-1}\left( F\left[ P%
\right] +\frac{\hbar ^{2}}{2m}\partial _{x}^{2}\right) \Psi ^{\ast } 
\]
an equation locally valid. Since $\Psi $\ and $\Psi ^{\ast }$ are
independent we have $F\left[ P\right] =-\frac{\hbar ^{2}}{2m}\partial
_{x}^{2}$ and ${\cal H}=-\frac{\hbar ^{2}}{2m}\partial _{x}^{2}+V$ or 
\begin{equation}
i\hbar \partial _{t}\Psi \left( x,t\right) =\left[ -\frac{\hbar ^{2}}{2m}%
\partial _{x}^{2}+V\left( x,t\right) \right] \Psi \left( x,t\right)
\label{schro}
\end{equation}
We have found that field equation for the probability amplitude $\Psi \left(
x,t\right) $\ follows quantum dynamics although the formalism used in this
work be classical. This classical signature of Stochastic Mechanics as
presented here is not only conceptual but has rigorous formal support as
shown below.

The whole perturbative series can now be recursively calculated. Expected
values of any (classical) observable $f\left( x,p,t\right) $ is given by 
\begin{equation}
\left\langle f\right\rangle =\sum_{n}\int a_{n}\left( x,t\right) \widetilde{f%
}\left( x,y,t\right) ^{\ast }y^{n}\exp \left( \frac{-y^{2}}{2\lambda ^{2}}%
\right) dxdy  \label{expec value}
\end{equation}
where $\widetilde{f}\left( x,y,t\right) =\int f\left( x,p\right) e^{-i\frac{%
py}{\hbar }}$. This way all predictions supported on the classical realism
can be obtained although satisfying quantum dynamics exactly as present on
the first two coefficients of the perturbative series. Higher order terms
are also influenced by quantum dynamics but it is not clear that in the full
phase space reconstruction interferences cancel out in order to recover the
classical character of the theory. If this was the case averages calculated
by eqn$\left( \ref{expec value}\right) $\ will result on the values expected
by classical theories. However it is noticeable that this doesn't happens
and concerning eqn$\left( \ref{expec value}\right) $\ some results from
Quantum Theory are equivalent to the present formulation of the Stochastic
Mechanics. For example the expected value of the (classical) Hamiltonian is,
using eqns$\left( \ref{expec value}\right) $ and$\left( \ref{a2}\right) $%
\begin{eqnarray*}
\left\langle H\right\rangle &=&\left\langle \frac{p^{2}}{2m}+V\left(
x,t\right) \right\rangle =\frac{-\hbar ^{2}}{m}\int a_{2}\left( x,t\right)
dx+ \\
&&\frac{\hbar ^{2}}{2m\lambda ^{2}}\int a_{0}\left( x,t\right) dx+\int
V\left( x,t\right) a_{0}\left( x,t\right) dx
\end{eqnarray*}
The value of $a_{2}$ is obtained by direct integration of eqn$\left( \ref{a1}%
\right) $ and use of the field equation for $\Psi $, the Shr\"{o}edinger
equation. The result is 
\begin{equation}
a_{2}\left( x,t\right) =\frac{1}{8}\left[ \Psi \frac{\partial ^{2}\Psi
^{\ast }}{\partial x^{2}}+\Psi ^{\ast }\frac{\partial ^{2}\Psi }{\partial
x^{2}}-2\left| \frac{\partial \Psi }{\partial x}\right| ^{2}\right] +\frac{%
a_{0}\left( x,t\right) }{2\lambda ^{2}}  \label{a22}
\end{equation}
Using the condition that norm of the probability amplitude is equal to one
the integral of \ eqn $\left( \ref{a22}\right) $\ is 
\[
\frac{-\hbar ^{2}}{m}\int a_{2}\left( x,t\right) dx=\frac{-\hbar ^{2}}{2m}%
\int \Psi ^{\ast }\frac{\partial ^{2}\Psi }{\partial x^{2}}dx-\frac{\hbar
^{2}}{2m\lambda ^{2}} 
\]
resulting that $\left\langle H\right\rangle =\left\langle H\right\rangle
_{MQ}$\ exactly, where $\left\langle H\right\rangle _{MQ}$\ is the expected
value of the Hamiltonian calculated by Quantum Mechanics for the same
problem. In particular for stationary states of bound systems the classical
energy becomes quantized. Consequently in the context presented here it
appears that Stochastic Mechanics is a larger class of dynamical problem
than Quantum Mechanics which seems be a subset of the former. In fact
Stochastic Mechanics gives a maximum possible information in phase space
through a true independence between position and momentum variables that is
not present in Quantum Theory.

But the above results show also that we have a disagreement between both
theories. Remember that Wigner function in phase space satisfies Liouville
equation only approximately \cite{balantine} whereas our probability density
satisfy it exactly for zero noise power. More generally Wigner function for
pure states in the adjoint space is given by $\widetilde{W}\left(
x,y,t\right) =\Psi \left( x-\frac{1}{2}y,t\right) \Psi \left( x+\frac{1}{2}%
y,t\right) ^{\ast }$\ and admits a series expansion of the form 
\[
\widetilde{W}\left( x,y,t\right) =\sum_{n=0}^{\infty }a_{wn}\left(
x,t\right) y^{n} 
\]
The first two coefficients of this series coincide with ours given by eqns $%
\left( \ref{def a0}\right) $\ and $\left( \ref{def a1}\right) $ but the
value of the second order coefficient given by 
\[
a_{w2}\left( x,t\right) =\frac{1}{4}\left[ \Psi \left( x,t\right) \frac{%
\partial ^{2}\Psi \left( x,t\right) ^{\ast }}{\partial x^{2}}+\Psi ^{\ast
}\left( x,t\right) \frac{\partial ^{2}\Psi \left( x,t\right) }{\partial x^{2}%
}-\left| \frac{\partial \Psi \left( x,t\right) }{\partial x}\right| ^{2}%
\right] 
\]
differ from the corresponding coefficient $a_{2}\left( x,t\right) $\
calculated in this work, even for infinite vacuum wavelength, and shown in
eqn $\left( \ref{a22}\right) $. This conclusively shows that Wigner function 
$W\left( x,p,t\right) $\ and $\Phi \left( x,p,t\right) $\ are not equal.

Having the terms in the perturbative series next step is phase space
reconstruction. This task is not trivial because involves the whole series
above displayed but if momentum of the mechanical system under study is much
larger than $p_{vac}\equiv \hbar /\lambda $ the gaussian weight may induce a
strong convergence in the perturbative series and truncation has some sense.
Thus we expect rapid convergence for sufficiently high energy of the
mechanical system compared to the mean vacuum fluctuation energy ${\cal E}$.
In this case the result up to second order is 
\begin{align}
\Phi \left( x,p,t\right) & \simeq G\left( x,p,t\right) \exp \left( -\frac{1}{%
2}\left( \frac{\lambda p}{\hbar }\right) ^{2}\right)  \label{phase space} \\
G\left( x,p,t\right) & \equiv a_{0}\left( x,t\right) \sqrt{2\pi \lambda }-%
\frac{i\lambda ^{3}p}{2\hbar }a_{1}\left( x,t\right) \sqrt{\pi }+  \nonumber
\\
& \frac{\lambda ^{3}}{3\hbar }a_{2}\left( x,t\right) \sqrt{\frac{2\pi }{3}}%
\left[ 1-\frac{1}{3}\left( \frac{\lambda p}{\hbar }\right) ^{2}\right] 
\nonumber
\end{align}
The above result shows that as an approximation, negative phase space
probability also exist in our theory as it happens with Wigner function in
Quantum Mechanics. Of course exact probability distribution should be
non-negative and depends on calculation of all terms in the series for $\chi 
$. It may happens that covergence be not uniform a fact that complicate
experimental verification since an oscilating series presents intrinsic
measurements dificulties. Interestingly more, a new effect rises out from eqn%
$\left( \ref{phase space}\right) $ as shown in its exponential prefactor
where a low energy phenomena controll a high energy cutoff in phase space,
being this a solution somewhat different from those usually found to
eliminate ultraviolet divergences in field theory. Even though of
fundamental importance this effect will not considered here for further
discussions.

\subsection{Concluding Remarks}

In this work a non-dissipative Brownian motion was considered and a
perturbative series for the associated Fokker-Planck equation presented.
Noise source was assumed to be vacuum fluctuations of some quantum field
that couples to particle's classical dynamics. Surprisingly the first two
terms of this series accommodates quantum dynamics quite naturally with
properties and structure of its Hilbert space stuff. Nevertheless our
approach is purely classical, explicit in the physical meaning of all
mathematical objects treated here. Moreover our probability density function
and Wigner function differ suggesting that both theories can be compared at
a experimental level specially if phase space tomography measurements \cite
{smithey} with massive particles becomes effectiveness.

In one line of reasoning this may indicate that our approach do not describe
a truly quantal phenomena since a disagreement with Wigner function occur.
But an alternative way may be given. Wigner function was originally
constructed as a mathematical object without any explicit classical
equivalence in what concern its (classically-defined) measurement. Its
simultaneous dependence on position and momentum variables indicates it
lives naturally in phase space. In addition it also satisfies approximately
Liouville equation but since it can present negative values no support
exists for an interpretation it is the quantum version of phase space
probability density function.

Quite differently, in the formalism presented here phase space dynamics is
real by construction and the interpretation of all mathematical objects are
philosophically self-consistent since they satisfy all paradigms of
Classical Mechanics reasoning. Even the probability waves and the associated
wavelengths $y$\ have clear physical meaning. In addition phase space
reconstruction shown in eqn$\left( \ref{phase space}\right) $ presents in
its lower order expansion negative probability values thus disclaiming the
interpretation where this phenomenon is a mark of the Quantum Theory. In
fact it is known from long that small noise perturbation series of the FPE
may present negative terms \cite{gardiner}. These considerations demands
that experiment should definitively test whether dynamics in phase space as
presented in this work is correct or not.

The most serious objection against the formulation of Stochastic Mechanics
we present is probably the possibility of a material body gain some energy
from vacuum. This reasoning breaks the naive classical paradigms we claim to
give to the rest of the theory. It is the price paid for get so smoothly
quantum dynamics from a purely classical theory. We have argued that
although vacuum power seems unphysical at first sight may in fact be a real
phenomenon with true detection chance. If real is reasonable think that
dynamcs of elementary particles are then correctly described by an
appropriate FPE as discussed here. Vacuum temperature must be so small that
usual quantum phenomena behave like high temperature ones and this in our
interpretation means that typical wavelengths of the probability density are
much smaller than vacuum wavelength so the whole perturbation series can be
approximated by its stationary Liouville solution. Consequently important
quantum results like atomic spectra can be understood inside our classical
reasoning. Moreover our formalism open tips to think about experimental
detection of alternative theories that not only embraces Quantum Mechanics
as well test its foundations at a more precise level. These considerations
are, in our assessment, sufficient to take account in serious way the
possibility that non-dissipative Brownian motion driven by vacuum
fluctuations be a real phenomena turning out Stochastic Mechanics a faithful
description of Nature.

\end{document}